# Spectrophotometric properties of dwarf planet Ceres from the VIR spectrometer on board the Dawn mission

M. Ciarniello[1], M. C. De Sanctis[1], E. Ammannito[2], A. Raponi[1], A. Longobardo[1], E. Palomba[1], F. G. Carrozzo[1], F. Tosi[1], J.-Y. Li[3], S. E. Schröder[4], F. Zambon[1], A. Frigeri[1], S. Fonte[1], M. Giardino[1], C. M. Pieters[5], C. A. Raymond[6], and C. T. Russell[2]

[1] IAPS-INAF, via Fosso del Cavaliere, 100, 00133 Rome, Italy
[2] University of California Los Angeles, Earth Planetary and Space Sciences, Los Angeles, CA, USA
[3] Planetary Science Institute, Tuscon, AZ 85719, USA
[4] German Aerospace Center DLR, Institute of Planetary Research, 12489 Berlin, Germany
[5] Department of Earth, Environmental, and Planetary Sciences, Brown University, Providence, RI, USA
[6] Jet Propulsion Laboratory, California Institute of Technology, Pasadena, USA

**ABSTRACT**

*Aims.* We present a study of the spectrophotometric properties of dwarf planet Ceres in the visual-to-infrared (VIS-IR) spectral range by means of hyper-spectral images acquired by the VIR imaging spectrometer on board the NASA Dawn mission.
*Methods.* Disk-resolved observations with a phase angle within the $7° < \alpha < 132°$ interval were used to characterize Ceres' phase curve in the 0.465–4.05 $\mu$m spectral range. Hapke's model was applied to perform the photometric correction of the dataset to standard observation geometry at VIS-IR wavelength, allowing us to produce albedo and color maps of the surface. The *V*-band magnitude phase function of Ceres has been computed from disk-resolved images and fitted with both the classical linear model and H-G formalism.
*Results.* The single-scattering albedo and the asymmetry parameter at 0.55 $\mu$m are $w = 0.14 \pm 0.02$ and $\xi = -0.11 \pm 0.08$, respectively (two-lobe Henyey-Greenstein phase function); at the same wavelength, Ceres' geometric albedo as derived from our modeling is $0.094 \pm 0.007$; the roughness parameter is $\bar{\theta} = 29° \pm 6°$. Albedo maps indicate small variability on a global scale with an average reflectance at standard geometry of $0.034 \pm 0.003$. Nonetheless, isolated areas such as the Occator bright spots, Haulani, and Oxo show an albedo much higher than average. We measure a significant spectral phase reddening, and the average spectral slope of Ceres' surface after photometric correction is 1.1% kÅ$^{-1}$ and 0.85% kÅ$^{-1}$ at VIS and IR wavelengths, respectively. Broadband color indices are $V-R = 0.38 \pm 0.01$ and $R-I = 0.33 \pm 0.02$. Color maps show that the brightest features typically exhibit smaller slopes. The H-G modeling of the *V*-band magnitude phase curve for $\alpha < 30°$ gives $H = 3.14 \pm 0.04$ and $G = 0.10 \pm 0.04$, while the classical linear model provides $V(1, 1, 0°) = 3.48 \pm 0.03$ and $\beta = 0.036 \pm 0.002$. The comparison of our results with spectrophotometric properties of other minor bodies indicates that Ceres has a less back-scattering phase function and a slightly higher albedo than comets and C-type objects. However, the latter represents the closest match in the usual asteroid taxonomy.

## 1. Introduction and rationale

After departing from asteroid Vesta in July 2012, the NASA Dawn spacecraft entered orbit around the dwarf planet Ceres on 6 March 2015. Before arrival, payload instruments started acquiring remote sensing data of Ceres from a distance of $1.2 \times 10^6$ km on 1 December 2014 during the early approach phase, making this body the first dwarf planet and the thirteenth asteroid-like object to be imaged by a space-based mission, after Gaspra (Helfenstein et al. 1994), Ida, and Dactyl (Helfenstein et al. 1996) from the Galileo spacecraft; Mathilde (Clark et al. 1999) and Eros (Clark et al. 2002; Li et al. 2004) from the Near Shoemaker mission; Braille (Buratti et al. 2004) from the Deep Space mission; Annefrank (Hillier et al. 2011) from Stardust; Itokawa (Fujiwara et al. 2006) from Hayabusa; Steins (Tosi et al. 2010; Spjuth et al. 2012) and Lutetia (Tosi et al. 2012; Schröder et al. 2015; Longobardo et al. 2016) from Rosetta; Toutatis (Zou et al. 2014) from Chang'e 2; and Vesta from DAWN (Li et al. 2013b; Schröder et al. 2013; Longobardo et al. 2014).

Ceres is the largest body in the main asteroid belt, located at an average heliocentric distance of ≈2.77 AU, with a mass of $9.38 \times 10^{20}$ kg and an average radius of ≈470 km (Park et al. 2016). Interior models based on shape data indicated that Ceres is a gravitationally relaxed and likely differentiated object (Thomas et al. 2005; Castillo-Rogez & McCord 2010) with a rocky core and a hydrated (icy or liquid) mantle (Neveu et al. 2015; Neveu & Desch 2015). Recent observations from Dawn and gravity data (Park et al. 2016) point instead to a high-density outer shell dominated by rocky material (Castillo-Rogez et al. 2016). Spectral modeling of surface spectra acquired by the Dawn mission shows widespread ammoniated phyllosilicates (De Sanctis et al. 2015; Ammannito et al. 2016) and abundant carbonate deposits in localized areas due to aqueous alteration (De Sanctis et al. 2016) as well as small spots of water ice in the northern hemisphere (Combe et al. 2016).

The Dawn spacecraft (Russell & Raymond 2011) carries three instruments: the Framing Camera (FC, Sierks et al. 2011), the Visible and Infrared spectrometer (VIR, De Sanctis et al. 2011), and the Gamma Ray and Neutron Detector (GRAND, Prettyman et al. 2011). VIR, whose dataset is investigated in this work, is an imaging spectrometer, working in the 0.2–5.1 $\mu$m spectral interval, using two channels covering the visible (VIS, 0.25–1.05 $\mu$m) and infrared (IR, 1–5.1 $\mu$m) ranges, with a spectral sampling of 1.8 nm and 9.5 nm, respectively, and an instantaneous field of view (IFOV) of 250 $\mu$rad × 250 $\mu$rad. The output of the instrument are hyper-spectral images ("cubes") that allow characterizing the spatial and spectral distribution of the light reflected from the target as well as measuring its thermal emission by exploring the long wavelength range 4.5–5.1 $\mu$m (Tosi et al. 2014).

The study of the photometric properties of airless bodies by means of remote sensing observations at different wavelengths allows simultaneously investigating the physical properties of their surface and composition. Furthermore, the derivation of a photometric model from a set of observations of a given target permits us to perform a photometric correction of the dataset itself (Li et al. 2015; Ciarniello et al. 2015), thus decoupling intrinsic surface albedo variability from effects related to the observing geometry. To achieve this twofold aim, we take advantage of VIR observations acquired from December 2014 up to June 2015 to study the global spectrophotometric properties of the Ceres surface.

After a description of the dataset in Sect. 2, we describe the derivation of the photometric correction of VIR observations by means of Hapke's model (Hapke 2012; Sect. 3), assessing the average spectrophotometric properties of the surface. Photometrically corrected data are then used to produce albedo maps and red, green, and blue (RGB) color maps in both the VIS and IR ranges (Sect. 4). Finally we compare the derived photometric properties of Ceres with similar measurements reported in previous works and with those derived for other minor bodies (asteroids and comets, Sect. 5), while a summary of the main findings of this paper is provided in Sect. 6.

## 2. Dataset

The VIR observations discussed in this study are from four different mission phases: Ceres Approach (CSA), Rotational Characterization 3 (RC3), Ceres Transfer to Survey (CTS), and Ceres Survey (CSS). Different mission phases are characterized by different spacecraft-target distances (Fig. 1), which progressively reduce from CSA (maximum altitude over the surface 359 006 km) to CSS (minimum altitude over the surface 4390 km). According to this, the spatial resolution at ground varies from ≈90 km/pix to ≈1 km/pix, as shown in Fig. 2. This dataset enables a proper characterization of Ceres' spectrophotometric properties at global scale, which is the focus of this work. For this reason, the High Altitude Mapping Orbit (HAMO, distance 1450 km) and Low Altitude Mapping Orbit (LAMO, distance 375 km), which are characterized by a higher spatial resolution and complete the Dawn mapping sequences, have not been analyzed in this study. Along with the spacecraft-target distance, the solar phase angle $\alpha$ also varies across different observation sequences, covering the overall 7° < $\alpha$ < 132° range (Fig. 1). The whole dataset we investigated counts a total of 305 VIS-IR hyper-spectral images, which corresponds to more than $3.6 \times 10^6$ spectra. These spectra have been corrected for spectral artifacts (Carrozzo et al. 2016; Raponi 2015) and thermal emission (Raponi et al. 2017). Given this, the available spec-tral range spans from 0.465 $\mu$m to 4.05 $\mu$m, corresponding to the interval where a high signal-to-noise ratio (S/N) is achieved. This prominent dataset allows us to characterize a large portion of the Ceres (≈90%), and, providing multiple observations of the different regions under different observation geometries, gives us the possibility to constrain the photometric behavior of the surface.

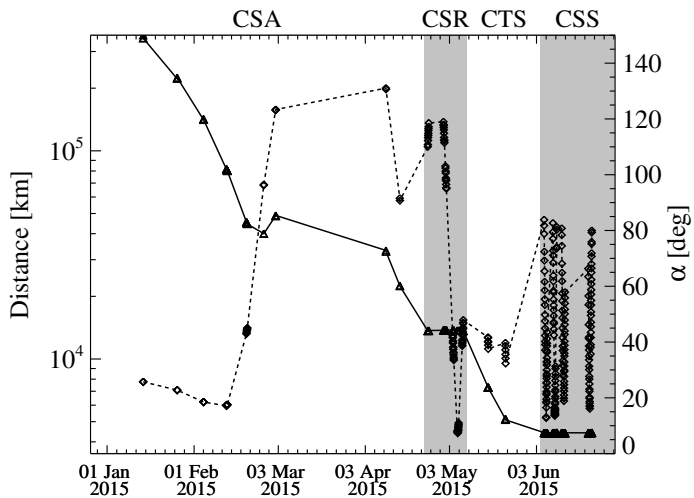

**Fig. 1.** Average target-spacecraft distance (solid line with triangles) and phase angle (dashed line with diamonds) for the VIR acquisitions investigated in this work. Each symbol corresponds to one observation. The different orbit phases are indicated.

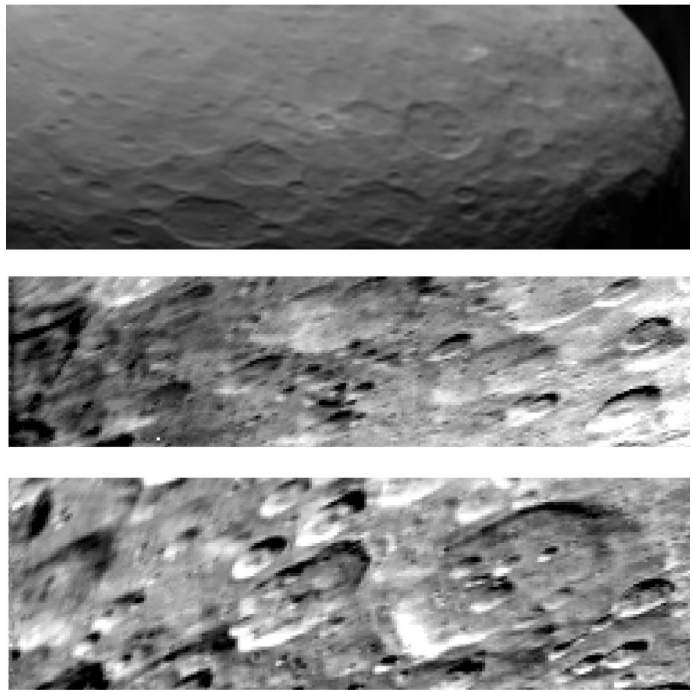

**Fig. 2.** Ceres surface as observed by VIR. The three different images at 0.55 $\mu$m are taken during RC3 (*top panel*, resolution of 3.4 km/pix), CTS (*central panel*, resolution of 1.3 km/pix), and CSS (*bottom panel*, resolution of 1.1 km/pix) phases.

## 3. Photometric correction

The spectrophotometric output of a particulate medium in a remote sensing experiment depends on the physical and compositional properties of the target as well as on the observation geometry, which is represented by the incidence $i$, emission $e$, and phase $\alpha$ angles. To compare images taken under different observation geometries, it is therefore necessary to correct them for photometrical effects to exploit the intrinsic variability related to the properties of the medium. This can be done by applying photometric models that describe the measured bidirectional reflectance $r$ (or radiance factor $I/F = \pi r$) as a function of $i$, $e$, and $\alpha$. Such photometric models can be either empirical (Minnaert 1941; Akimov 1988), semi-empirical (Longobardo et al. 2014), or physically based as in the case of the Shkuratov et al. (1999) model and Hapke's theory (Hapke 2012). The latter has been widely used in spectrophotometric analysis of planetary objects, as shown in Domingue & Verbiscer (1997), Domingue et al. (2010), Ciarniello et al. (2011), Li et al. (2013a), Déau (2015), and (Ciarniello et al. 2015), and in particular for asteroids (Clark et al. 1999, 2002; Domingue et al. 2002; Li et al. 2004, 2006; Reddy et al. 2015). In this work we applied the Hapke model to perform a photometric correction of VIR dataset following an approach similar to Ciarniello et al. (2015) with three main differences: adoption of a two-parameter Henyey-Greenstein single-particle phase function (SPPF; Hapke 2012), instead of the single-term expression (Henyey & Greenstein 1941); inclusion of the shadow-hiding opposition effect (SHOE); inclusion of the multiple scattering term. Given this, the expression of the bidirectional reflectance we use to perform the photometric correction is

$$r(i,e,\alpha) = \frac{w}{4\pi} \frac{\mu_{0\mathrm{eff}}}{\mu_{0\mathrm{eff}} + \mu_{\mathrm{eff}}}$$
$$\times [B_{\mathrm{SH}}(\alpha)p(\alpha) + H(w,\mu_{0\mathrm{eff}})H(w,\mu_{\mathrm{eff}}) - 1]$$
$$\times S(i,e,\alpha,\bar{\theta}), \qquad (1)$$

where $w$ is single-scattering albedo (SSA); $\mu_{0\mathrm{eff}}$ and $\mu_{\mathrm{eff}}$ are the effective cosines of the incidence and emission angles, respectively; $B_{\mathrm{SH}}(\alpha) = \frac{B_0}{1+1/h\tan(\alpha/2)}$ is the SHOE term depending on the opposition surge angular width $h$ and amplitude $B_0$; $p(\alpha) = \frac{1+c}{2}\frac{1-b^2}{(1-2b\cos(\alpha)+\alpha^2)^{3/2}} + \frac{1-c}{2}\frac{1-b^2}{(1+2b\cos(\alpha)+\alpha^2)^{3/2}}$ is the two-parameter SPPF where $b$ ranges in the $[0;1)$ interval and represents the width of the SPPF back-scattering and forward-scattering lobes and $c$ determines the SPPF behavior: back-scattering ($c>0$) or forward-scattering ($c<0$); $H(w,x)$ is the Chandrasekhar function (Chandrasekhar 1960), accounting for the multiple scattering contribution; $S(i,e,\alpha,\bar{\theta})$ is the shadowing function depending on the surface average slope parameter $\bar{\theta}$.

With respect to the complete version of the Hapke model, this expression does not include the coherent backscattering opposition effect (CBOE) (Hapke 2002) and does not take into account the effects of porosity described in Hapke (2008) and Ciarniello et al. (2014). This is justified because CBOE is significant at very small phase angles ($<2°$) that are beyond the range investigated in this work, while as discussed in Li et al. (2013a) and Ciarniello et al. (2015), the effect of porosity cannot be decoupled by SSA for low-albedo objects such as Ceres. According to this, the free parameters in the model are $w, b, v, B_0,$ and $h$, all depending on the wavelength, and $\bar{\theta}$, which, being related to surface morphology, has no spectral dependence. Since the smallest investigated phase angle is $7°$, it is not possible to model the opposition surge and constrain the SHOE parameters $B_0, h$. We therefore assumed the values reported by Helfenstein & Veverka (1989) throughout the whole investigated spectral range, $B_0 = 1.6$ and $h = 0.06$.

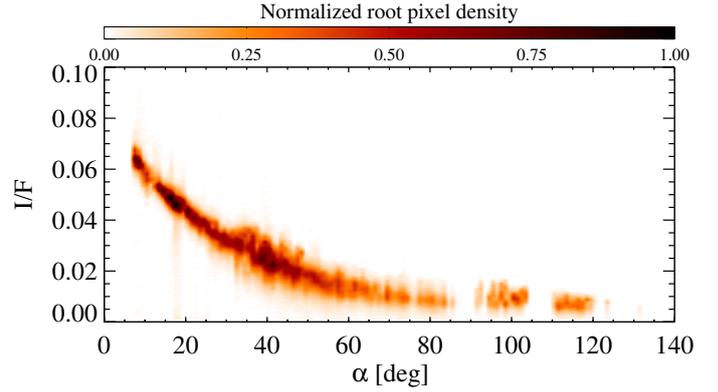

**Fig. 3.** $I/F$ at 0.55 $\mu$m as a function of the phase angle: contour plot showing the square root of point density normalized to its maximum value.

The whole dataset used to derive the photometric correction at 0.55 $\mu$m wavelength is shown in Fig. 3, where $I/F$ is reported as a function of the phase angle. Pixels acquired with large incidence and emission angle ($i, e > 70°$) and very low signal ($I/F < 0.001$ at 0.55 $\mu$m) have been filtered out. This allowed us to minimize the effect that is due to unfavorable geometries, which can be affected by relatively larger errors and limit the occurrence of pixels in total or partial shadow.

The algorithm used to derive the set of Hapke's parameters for the Ceres dataset is similar to the one adopted in Ciarniello et al. (2015), but presents some significant diversities, therefore we consider it useful to report the main steps here:

- The I/F at a given wavelength of each pixel is multiplied by the factor $4\frac{\mu_{0\mathrm{eff}}+\mu_{\mathrm{eff}}}{\mu_{0\mathrm{eff}}S(i,e,\alpha,\bar{\theta}_0)}$, where $\bar{\theta}_0$ is a given value of the roughness slope parameter. Assuming Eq. (1), this quantity, which we call here equigonal albedo $A_{\mathrm{eq}}$ (Fig. 4) in analogy with Akimov's photometric model (Shkuratov et al. 1999), corresponds to the following expression: $A_{\mathrm{eq}}(\alpha,i,g) = w[B_{\mathrm{SH}}(\alpha)p(\alpha) + H(w,\mu_{0\mathrm{eff}})H(w,\mu_{\mathrm{eff}}) - 1]$. We note that equigonal albedo in Akimov's theory does not depend on $i$ and $e$. Conversely, our definition formally depends on incidence and emission angle through the Chandrasekhar function. However, in the case of low-albedo surfaces such as that of Ceres, $H(w,x) \approx 1$ and the dependence on $i$ and $e$ is weak.
- The equigonal albedo is averaged on phase angle bins of $1°$ obtaining $\langle A_{\mathrm{eq}} \rangle_\alpha$ (Fig. 4).
- $\langle A_{\mathrm{eq}} \rangle_\alpha$ is then fitted with its corresponding averaged expression from the Hapke theory $w\langle B_{\mathrm{SH}}p \rangle_\alpha + w\langle H(w,\mu_{0\mathrm{eff}})H(w,\mu_{\mathrm{eff}}) - 1\rangle_\alpha$, which depends on the photometric parameters of the model and is computed on the observation geometries of the investigated dataset.
- The output of this algorithm is a set of photometric parameters at each wavelength for the given $\bar{\theta}_0$: this is repeated by varying $\bar{\theta}$ in the range $[0°;60°]$ at steps of $1°$ (Fig. 5).
- The whole dataset is photometrically corrected using all the derived sets of parameters separately (characterized by different values of $\bar{\theta}$). This is done by computing the photometrically corrected radiance factor $I/F_{\mathrm{std}}$, which is the measured $I/F(i,e,\alpha)$ reported to standard geometry ($i = 30°$, $e = 0°, \alpha = 30°$) by means of the following relation:

$$I/F_{\mathrm{std}} = \frac{I/F(i,e,\alpha)}{I/F_m(i,e,\alpha)} I/F_m(30°,0,30°), \qquad (2)$$

where the subscript $m$ indicates the modeled radiance factor.

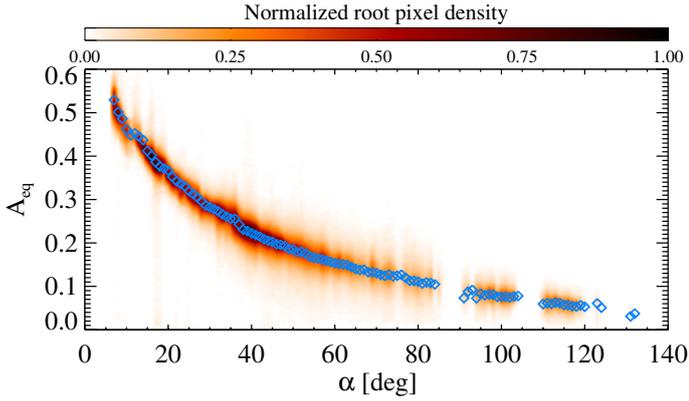

**Fig. 4.** $A_{eq}$ at 0.55 $\mu$m as a function of the phase angle for $\bar{\theta}_0 = 0°$: contour plot showing the square root of point density normalized to its maximum value. Blue diamonds represent the average equigonal albedo on 1° phase angle bins $\langle A_{eq} \rangle_\alpha$.

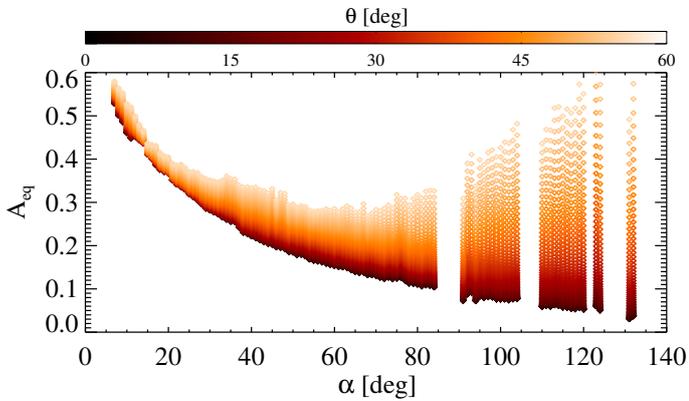

**Fig. 5.** Average $A_{eq}(\langle A_{eq} \rangle_\alpha)$ at 0.55 $\mu$m as a function of the phase angle for different $\bar{\theta}$. The increment of $\bar{\theta}$ produces an enlargement of $\langle A_{eq} \rangle_\alpha$ at large phase angles with respect to small angles as well as a slight increase of its overall level.

– The final set of parameters is defined as the one that at selected wavelengths (see Sect. 3.1) minimizes the residual correlation of $I/F_{std}$ with the observation geometry, that is, with the incidence, emission, and phase angles.

### 3.1. Surface roughness

The final surface roughness parameter is determined as the value that provides the minimal residual correlation of the corrected $I/F$ with incidence, emission, and phase angles. As stated above, this parameter is assumed to have no spectral dependence, being related to terrain morphology at scales much larger than the wavelength (Hapke 2012), and it is constrained using information from selected wavelengths in the VIS (0.545 $\mu$m $\leq \lambda \leq$ 0.946 $\mu$m) and in the IR (1.02 $\mu$m $\leq \lambda \leq$ 2.99 $\mu$m) that are characterized by good S/N and poorly affected by residual thermal emission. For a given $\bar{\theta}_0$ and the corresponding set of spectral parameters, we computed the corrected reflectance $I/F_{std}$ and performed averages on 1° bins of incidence, emission, and phase angles. This produced a set of three curves at each wavelength: $\langle I/F_{std} \rangle_i$, $\langle I/F_{std} \rangle_e$, and $\langle I/F_{std} \rangle_\alpha$. These quantities were then normalized at the smallest incidence, emission, and phase angle and were spectrally averaged, thus producing a final set for each value of $\bar{\theta}_0$, that we refer to as $C_i$, $C_e$, and $C_\alpha$ (Fig. 6). These curves quantify the residual correlation of the corrected $I/F$ with observation geometry for a given value of the slope parameter. Each set of $C_i$, $C_e$, and $C_\alpha$ was fitted with a line with null angular coefficient (it represents the ideal behavior in case of no residual correlation), and the resulting residuals of the three fits were summed to produce a $\chi^2$ variable, whose distribution as a function of the slope parameter is reported in Fig. 7. The final value of $\bar{\theta}$ was chosen as the value that produced the smallest $\chi^2$, from which we obtained $\bar{\theta} = 29° \pm 6°$. This value is much lower than the value derived in Li et al. (2006) $\bar{\theta} = 44°$ from HST data. This difference can be explained by the fact that in Li et al. (2006) a much smaller range of phase angles was explored (6.1°–6.2°), which prevented them from thoroughly constraining the effect of surface roughness, which is increasingly important at large $\alpha$. Another possible cause for the high value of $\bar{\theta}$ derived in Li et al. (2006) is that the point spread function of the high-resolution channel of the Advanced Camera for Surveys on HST, which was used to perform Ceres observations, was a few pixels wide, with the target being roughly 30 pixels across. This could produce a larger reduction of the signal toward the edge of Ceres' disk, resulting in a higher roughness parameter (J.-Y. Li, Priv. comm.). Recent evaluations of the roughness parameter from DAWN-FC observations report $\bar{\theta} = 20° \pm 3°$ (Li et al. 2016b) and $\bar{\theta} = 22° \pm 2°$ (Schröder et al. 2017). Our estimate of $\bar{\theta}$, although somewhat larger than the results of Li et al. (2016b) and Schröder et al. (2017), is still compatible with their values within the errors.

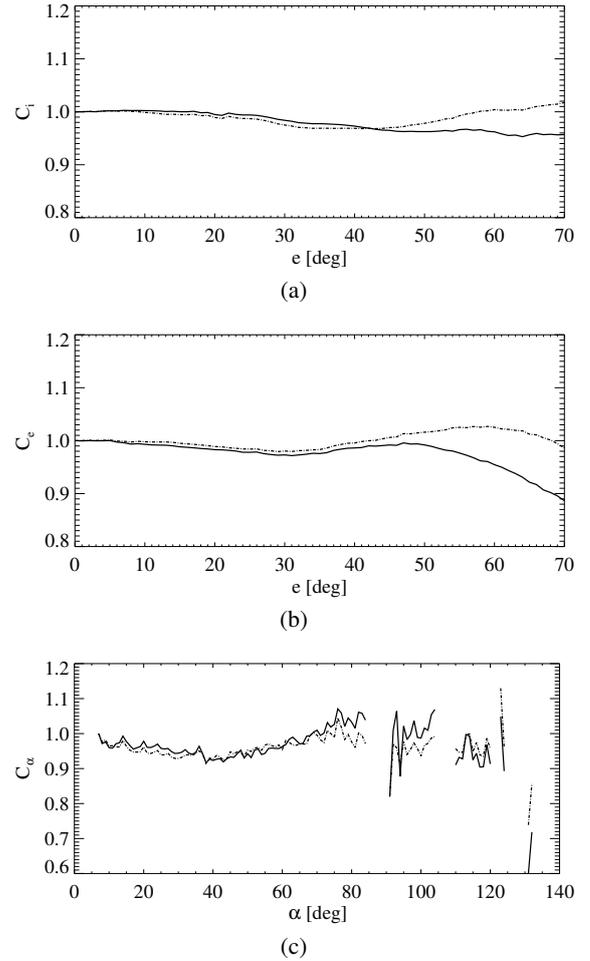

**Fig. 6.** $C_i$ (*top panel*), $C_e$ (*central panel*) and $C_\alpha$ (*bottom panel*). In each plot we show two curves for comparison, corresponding to $\bar{\theta}_0 = 0°$ (solid line) and $\bar{\theta}_0 = 29°$ (dot-dashed line). It can be noted that increasing the slope parameter allows us to compensate for the reduction of signal at large $i$ ($\approx$10%) and $e$ ($\approx$5%) while minimizing the amplitude of the modulation of the signal at large $\alpha$ for $C_\alpha$. A residual modulation of $C_\alpha$ with the phase angle is still present for $\bar{\theta}_0 = 29°$, possibly because regions with intrinsically different albedo have been observed at different phase angles or because of a deviation of the regolith SPPF behavior from the adopted Henyey-Greenstein formulation. Smaller residuals (<3%) can also be noted for $C_i$ and $C_e$.

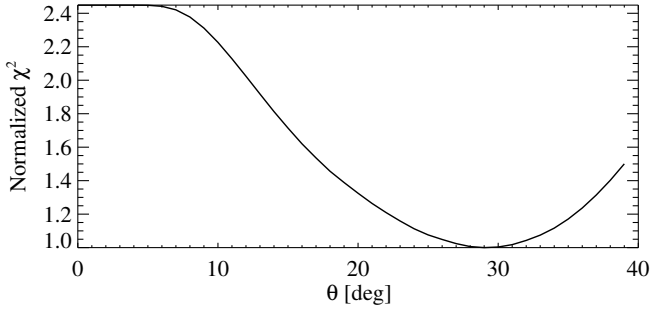

**Fig. 7.** $\chi^2$ normalized at its maximum value as a function of the slope parameter.

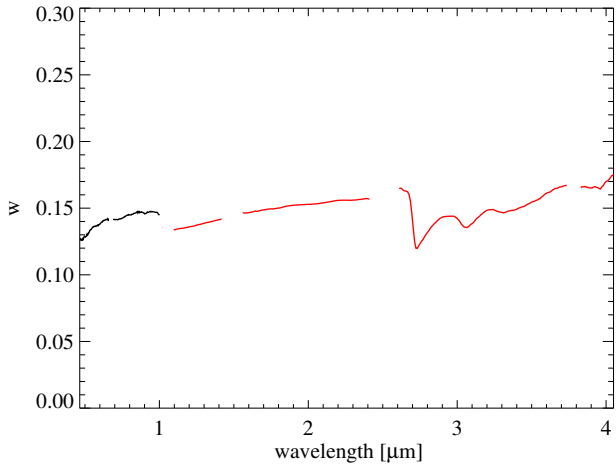

**Fig. 8.** SSA spectrum: VIS (black) and IR (red). Missing parts correspond to the order-sorting filter positions and the VIS-IR channels junction.

### 3.2. SSA

In Fig. 8 the derived single-scattering albedo is shown as a function of wavelength. This quantity can be considered as representative of the average spectral properties of the surface of Ceres, and it is linked to its composition. In the VIS range the SSA spectrum has a sharp absorption at the shortest wavelength with a moderate red slope from 0.47 $\mu$m up to 0.9 $\mu$m. The spectrum in the IR has a slightly positive slope from 1 $\mu$m up to 2.65 $\mu$m. Absorption features at 2.7 $\mu$m and 3.06 $\mu$m, attributed to -OH and -NH$_4$ bearing minerals, as well as carbonates bands at 3.3 $\mu$m and 3.95 $\mu$m can be recognized (De Sanctis et al. 2015). The value of the SSA at 0.55 $\mu$m is $w = 0.14 \pm 0.02$ (the dispersion is related to the error on $\bar{\theta}$ parameter), which is much larger than the value provided by Helfenstein & Veverka (1989; $w = 0.06$) and Li et al. (2006; $w = 0.07$) and different from the results derived for the various models described in Reddy et al. (2015; $w = 0.070, 0.11, 0.083$) and Li et al. (2016b; $w = 0.094-0.11$). It is important to mention that the analyses of Helfenstein & Veverka (1989), Li et al. (2006), Reddy et al. (2015) have been conducted on ground-based data, while Li et al. (2016b) used Dawn-FC observations in the 30°–90° phase angle range (effective wavelength 730 nm). All these works assume a single-parameter expression for the SPPF, given the relatively limited investigated phase angle range. The adoption of different versions of the Hapke model and the possibility of accessing a much larger portion of the phase curve with VIR data can account for the differences in the retrieved SSAs.

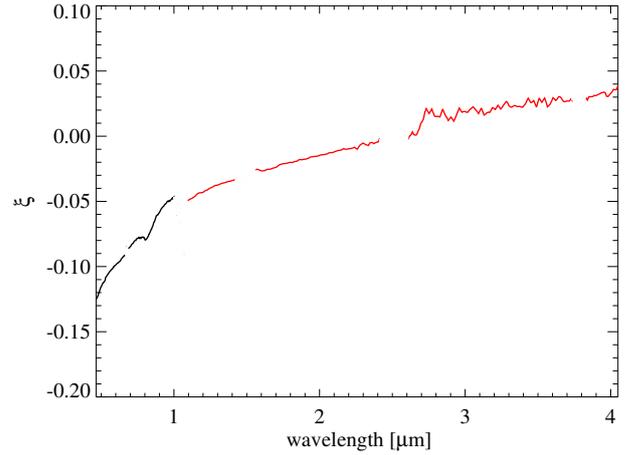

**Fig. 9.** $\xi$ as a function of wavelength: VIS (black) and IR (red). Missing parts correspond to the order-sorting filter positions and the VIS-IR channels junction.

### 3.3. SPPF and cosine asymmetry factor

As described in Sect. 3, the single-particle phase function depends on two parameters, $b$ and $v$, which determine the shape of the SPPF and its behavior as a function of the phase angle. These quantities can be combined in a single expression to give the cosine asymmetry factor $\xi = -bv = -\langle\cos(\alpha)\rangle$. It represents the opposite of the average value of the phase angle cosine weighted by the SPPF. Negative values of $\xi$ imply a a predominantly backscattering single-particle phase function that turns to forward-scattering for a positive asymmetry factor. A value of $\xi = 0$ indicates a symmetric SPPF. In Fig. 9 the spectral dependence of $\xi$ is shown as obtained from the derived values of $b$ and $v$. It can be noted that the behavior of the SPPF changes from backscattering at short wavelength to fairly symmetric in the IR, being partially correlated with the SSA. The value of the asymmetry factor at 0.55 $\mu$m is $\xi = -0.11 \pm 0.08$. The large dispersion associated to this quantity is related to the error on the roughness parameter $\bar{\theta}$ and is due to the coupling among the different terms of the Hapke model and in particular between the SPPF and the shadowing function $S(i, e, \alpha, \bar{\theta})$.

### 3.4. Spectral slopes and phase reddening

The dependence of the SPPF on wavelength implies that the spectral output depends on phase angle. This is shown in Fig. 10, where the spectral slope in the VIS (0.55–0.80 $\mu$m) and IR (1.2–2 $\mu$m) ranges as derived by VIR observations are reported as a function of $\alpha$. These quantities are defined as $S_{VIS} = \frac{I/F_{0.80\,\mu m} - I/F_{0.55\,\mu m}}{I/F_{0.55\,\mu m}(8-5.5\,k\text{Å})}$ and $S_{IR} = \frac{I/F_{2\,\mu m} - I/F_{1.2\,\mu m}}{I/F_{1.2\,\mu m}(20-12\,k\text{Å})}$, respectively.

Both $S_{VIS}$ and $S_{IR}$ show a positive fairly linear correlation with phase angle (Figs. 10a, b), similarly to what is observed by Li et al. (2016b), with an increment of $4.6 \times 10^{-2}\% \, k\text{Å}^{-1} \, \text{deg}^{-1}$ in the VIS and $1.5 \times 10^{-2}\% \, k\text{Å}^{-1} \, \text{deg}^{-1}$ in the IR, which is removed after the photometric correction (Figs. 10c, d), giving $S_{VIS}$ = 1.1% kÅ$^{-1}$ and $S_{IR}$ = 0.85% kÅ$^{-1}$. This behavior, indicated as "phase reddening", is common for asteroids (see Li et al. 2015, and references therein) and airless bodies in general, such as comets (Ciarniello et al. 2015), icy satellites (Cuzzi et al. 2002; Ciarniello et al. 2011; Filacchione et al. 2012), and planetary rings (Filacchione et al. 2014; Ciarniello et al. 2016). Moreover, it has been observed in laboratory measurements and light-scattering simulations from Schröder et al. (2014). From these studies it emerges that two typical behaviors of the spectral slopes with phase angle can be recognized: arch-shaped (Schröder et al. 2014) and monotonic. The first, associated to semi-transparent and relatively bright particles, is explained by the combined effect of multiple scattering and surface shadowing (Kaydash et al. 2010), while the second is shown by opaque and low-albedo ones and might be related to particle surface roughness. The latter is in good agreement with the monotonic phase reddening measured by VIR at Ceres, which is characterized by a dark surface.

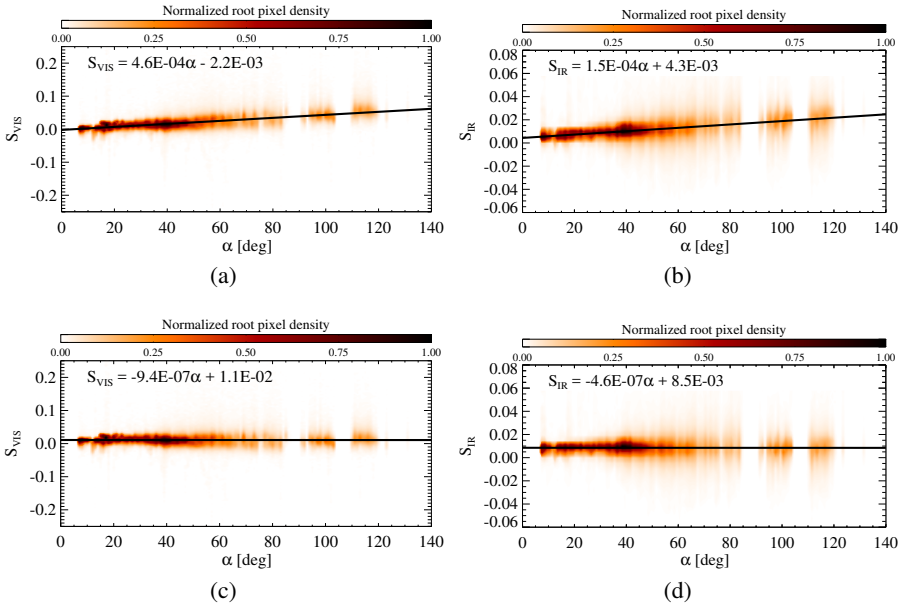

Fig. 10. Density plot of $S_{VIS}$ **a)** and $S_{IR}$ **b)** as a function of $\alpha$. The same quantities after photometric correction are reported in panels **c)** and **d)**. Linear fits to the distributions are reported as black solid lines in the plots, with their equations: the angular coefficient is expressed in [kÅ$^{-1}$ deg$^{-1}$] and intercept in [kÅ$^{-1}$].

### 3.5. Photometric correction accuracy

The derived set of photometric parameters is used in Eq. (2) to report the measured $I/F(i, e, \alpha)$ to its correspondent value in standard observation geometry $I/F_{std}$. This implies the assumption that surface photometric behavior is globally uniform and that changes in $I/F_{std}$ are due to the variability of the intrinsic albedo properties of the different observed regions. As shown below, this approximation is reasonable in the case of Ceres' surface, which shows small variability at large scale (>1 km). Nonetheless, a number of isolated bright features have been recognized throughout the surface of the body, as also reported by Schröder et al. (2017), De Sanctis et al. (2016), Russell et al. (2016). They show a much higher albedo than the average, which cannot be described by the same set of photometric parameters derived here. In these cases, the value of the $I/F_{std}$ is an approximated estimate of the real reflectance of the surface as observed in standard geometry.

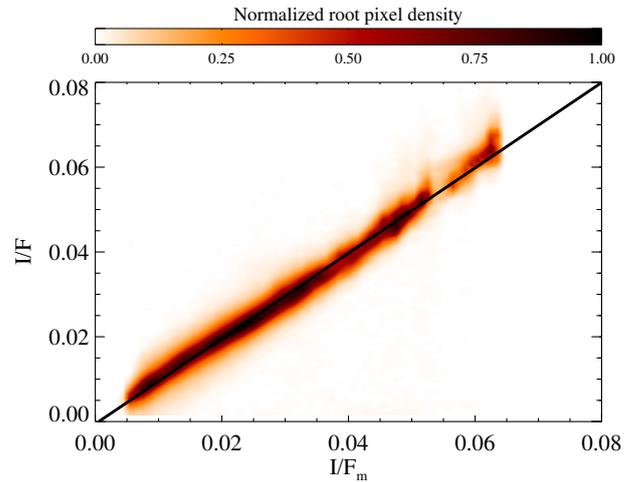

Fig. 11. $I/F$ against $I/F_m$ at 0.55 $\mu m$: the black solid line is the linear fit to the distribution with equation $I/F = 1.006 \times I/F_m - 5 \times 10^{-4}$.

To quantify the accuracy of the derived photometric correction, we report in Fig. 11 the measured $I/F$ at 0.55 $\mu m$ against the radiance factor at the same observation geometry as modeled with the derived Hapke $I/F_m$. The two quantities are linearly correlated ($R = 0.99$), as expected, and a linear fit of the distribution gives $I/F = 1.006 \times I/F_m - 5 \times 10^{-4}$, which is close to the ideal case $I/F = I/F_m$, while the average relative error between the modeled and measured reflectance is about 7%.

## 4. Albedo maps

To show the regional variability of surface brightness, photometrically corrected data $I/F_{std}$ were reprojected onto a latitude-longitude grid with a resolution of 0.25° × 0.25° to produce albedo maps of Ceres' surface. The projection was performed according to the real footprint of VIR pixels, and the median was calculated for overlapping pixels. With the aim to further minimize the effect of unfavorable geometries, only pixels with $i, e < 60°$ were selected. In Figs. 12 and 13 we report the albedo map at 0.55 $\mu m$ and 2 $\mu m$, respectively.

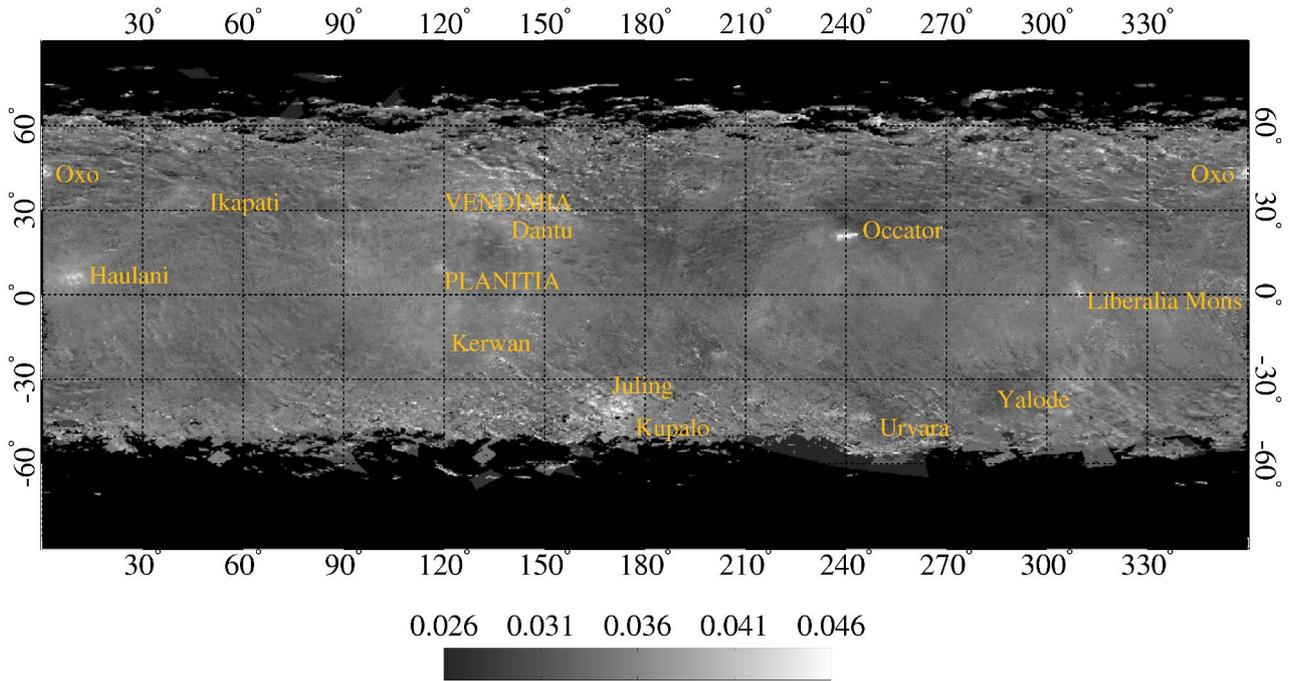

**Fig. 12.** Albedo map at 0.55 $\mu$m.

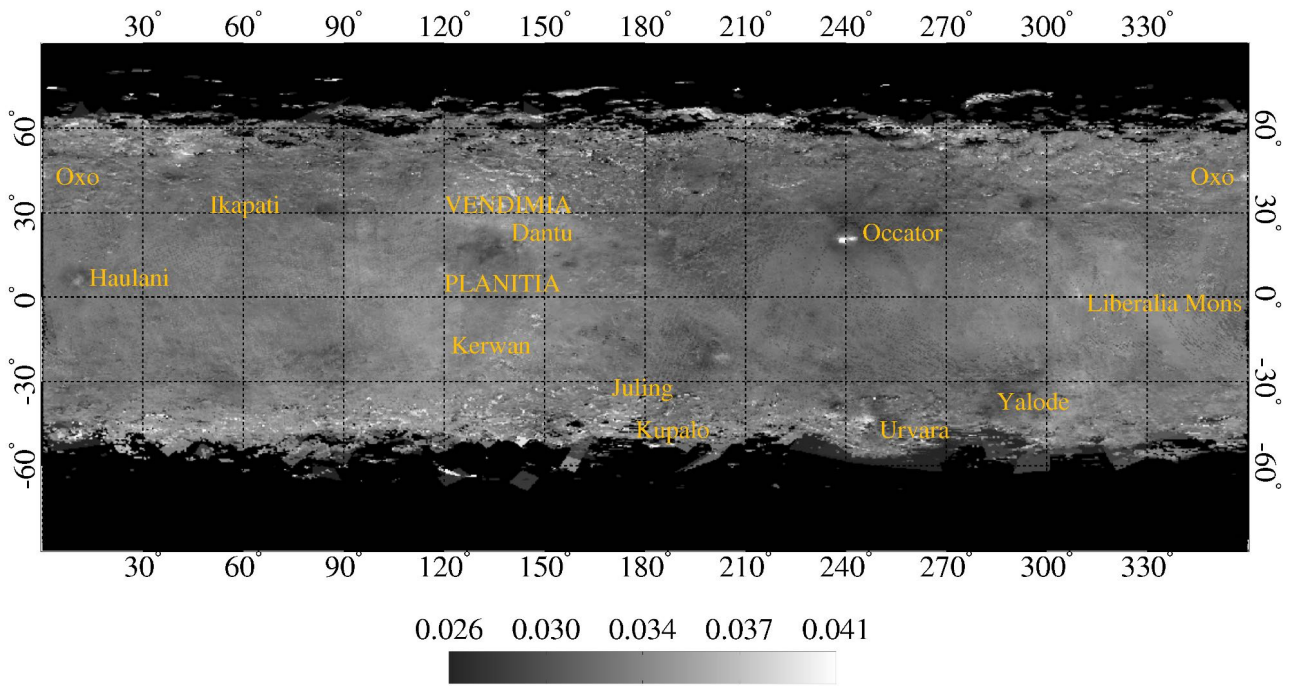

**Fig. 13.** Albedo map at 2 $\mu$m.

### 4.1. Albedo map at 0.55 $\mu$m

The albedo map at 0.55 $\mu$m (Fig. 12) indicates zonal reflectance variability. In particular, two adjacent extended areas with an albedo higher and lower than the average, respectively, are found in the equatorial region. Brighter terrains surround the central part of Vendimia Planitia region, with the largest albedo across and to the north of Dantu crater ($I/F_{\rm std} \approx 0.037$), in a region extending from $105° < {\rm lon} < 165°$ and $-30° < {\rm lat} < 45°$, which includes the Kerwan crater. East of this position, in the $165°-210°$ longitude interval, Ceres' surface appears darker than the average with $I/F_{\rm std} \approx 0.033$. The rest of the surface with $-30° < {\rm lat} < 30°$ exhibits an intermediate albedo $I/F \approx 0.034$. Isolated features corresponding to morphological structures can be recognized, as the Haulani (lon $\approx 11°$, lat $\approx 6°$, max $I/F_{\rm std} = 0.050$) and Oxo (lon $\approx 0°$, lat $\approx 42°$ max $I/F_{\rm std} = 0.068$) craters, the region corresponding to Juling and Kupalo (max $I/F \approx 0.057$) and the Liberalia Mons area ($I/F_{\rm std} = 0.041$).

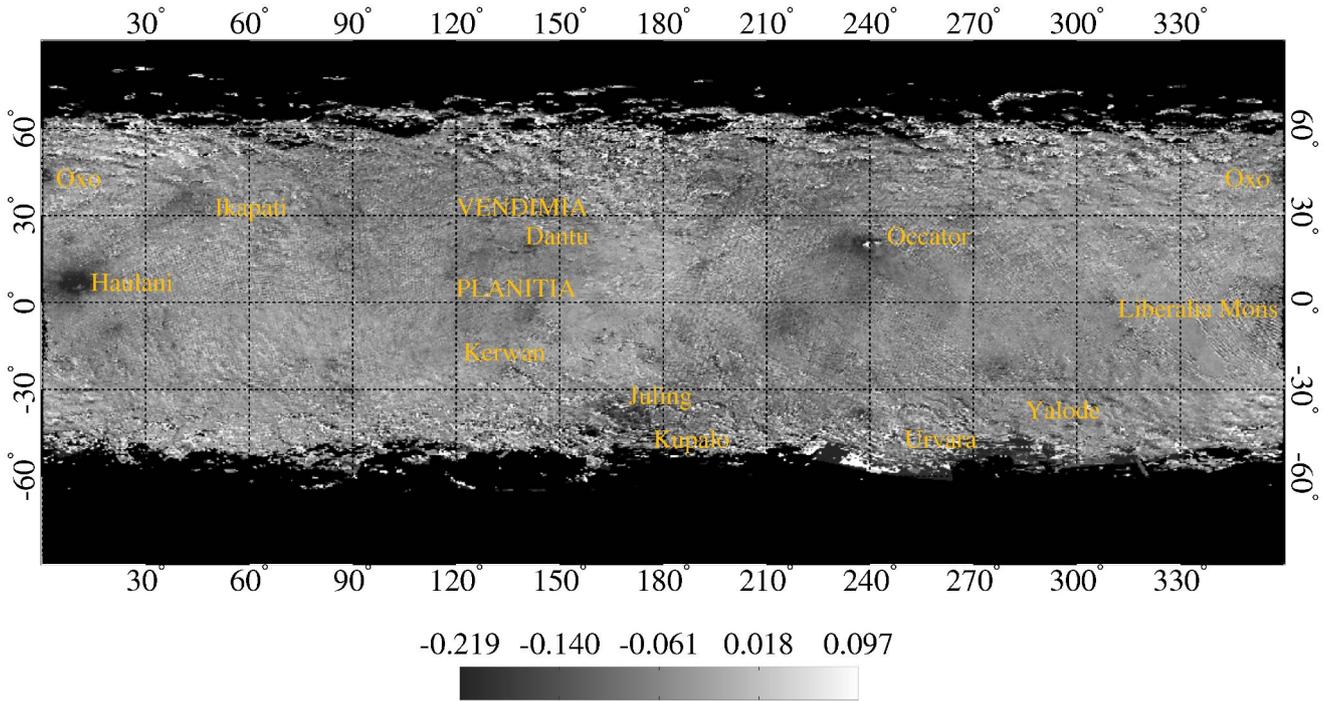

**Fig. 14.** Map of the albedo difference at 2 µm and 0.55 µm.

Bright spots in the Occator crater are not fully resolved because of the limited spatial resolution of the dataset. This region was observed mainly during the RC3 phase with a typical resolution of 3.4 km, from which we obtained a maximum value of $I/F_{std} = 0.12$, whereas De Sanctis et al. (2016) reported 0.26 from HAMO data. On the northeast face of the crater a portion characterized by low reflectance ($I/F_{std} \approx 0.031$) can be noted. In this case, the transition from high to low albedo is very sharp, and it is distributed along a linear feature extending from lon ≈ 232°, lat ≈ 32° to lon ≈ 252°, lat ≈ 14°. Two distinctly darker areas with respect to the nearby terrains are located corresponding to the Urvara and Yalode craters with $I/F_{std} \approx 0.032$. Albedo level tends to increase toward the poles, possibly as a result of overcorrecting the reflectance level because of unfavorable observation geometries (large $i$ and $e$). It is worth mentioning that the albedo distribution derived in this work is in good agreement with the distribution derived in Li et al. (2006) from HST images and the distribution obtained with observation from the Dawn Framing Camera (Schröder et al. 2017).

### 4.2. Albedo map at 2 µm

The albedo map at 2 µm (Fig. 13) appears to be fairly similar to the map derived at 0.55 µm, but it presents some peculiar differences. While albedo variability is preserved at large scale with minor modifications (see, for example, the shape of the bright region located at 105° < lon < 165° and −30° < lat < 45°), isolated structures in the VIS do not always show an IR counterpart. The most significant example is the Haulani crater, which is characterized by dark ejecta at 2 µm, showing an increase of brightness in the central part. This indicates that along with albedo variation, spectral variegation also occurs. To isolate regions with maximum spectral variability, we report in Fig. 14 the relative difference of the two maps, computed according to the following equation:

$$\Delta = 2\frac{I/F_{std}(2\ \mu m) - I/F_{std}(0.55\ \mu m)}{I/F_{std}(0.55\ \mu m) + I/F_{std}(2\ \mu m)}. \quad (3)$$

In addition to Haulani, Occator's floor also exhibits negative values of Δ, indicating spectral properties that are different from the rest of the surface. Bright spots inside Occator also show variability of the Δ parameter, with adjacent positive and negative values. Although De Sanctis et al. (2016) reported that these regions are rich in carbonates, the sharp variation of Δ shown here, given the limited extension of the spots, is more likely due to a slight displacement between the VIS and IR reprojected pixels than to compositional variability. Finally, the Oxo crater and a relatively larger region associated with the Juling and Kupalo craters also exhibits a negative Δ. All these areas, with the exception of the Occator floor, are characterized by high albedo in the VIS range, suggesting that small-scale high albedo features are characterized by a bluish spectrum. This is confirmed from the enhanced colors maps shown in Sect. 4.4.

### 4.3. Albedo variability

In Fig. 15 histograms for $I/F_{std}$ at 0.55 µm and 2 µm as derived from the corresponding albedo maps are shown. The distribution is unimodal in both cases, with average values of 0.034 ± 0.003 at 0.55 µm and 0.032 ± 0.002 at 2 µm. This indicates that on a global scale, the Ceres surface as observed at VIR resolution is fairly uniform. The width of the distributions of Fig. 15 is linked to the intrinsic albedo variability at the surface, to the error on the measured reflectance, and to the error associated with the photometric correction (uncertainties on observation geometry and photometric parameters). It is then possible to give an upper limit of the global albedo variability, which is on the order of 9% in the VIS and 6% in the IR. On the other hand, at a local scale, the albedo variability can be much higher, as in the case of the brightest features like the Occator bright spots, Oxo, and Haulani, which correspond to areas with limited extension.

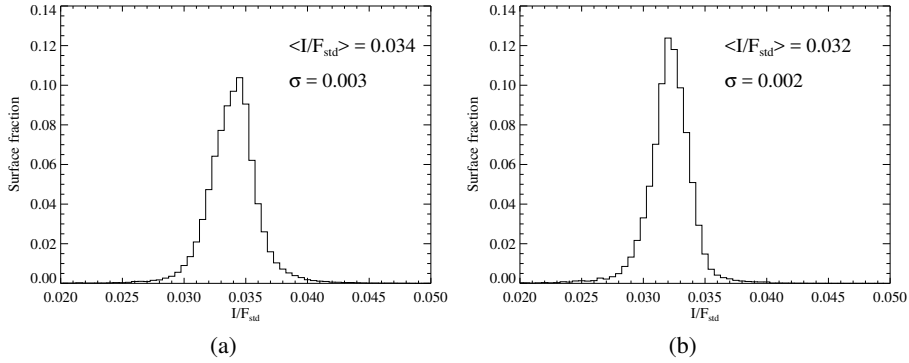

**Fig. 15.** Histograms of $I/F_{std}$ at 0.55 $\mu$m **a)** and 2 $\mu$m **b)**: the average values $\langle I/F_{std} \rangle$ and the standard deviation $\sigma$ of the distributions are indicated. Histograms are built according to the real area subtended by each pixel in the maps.

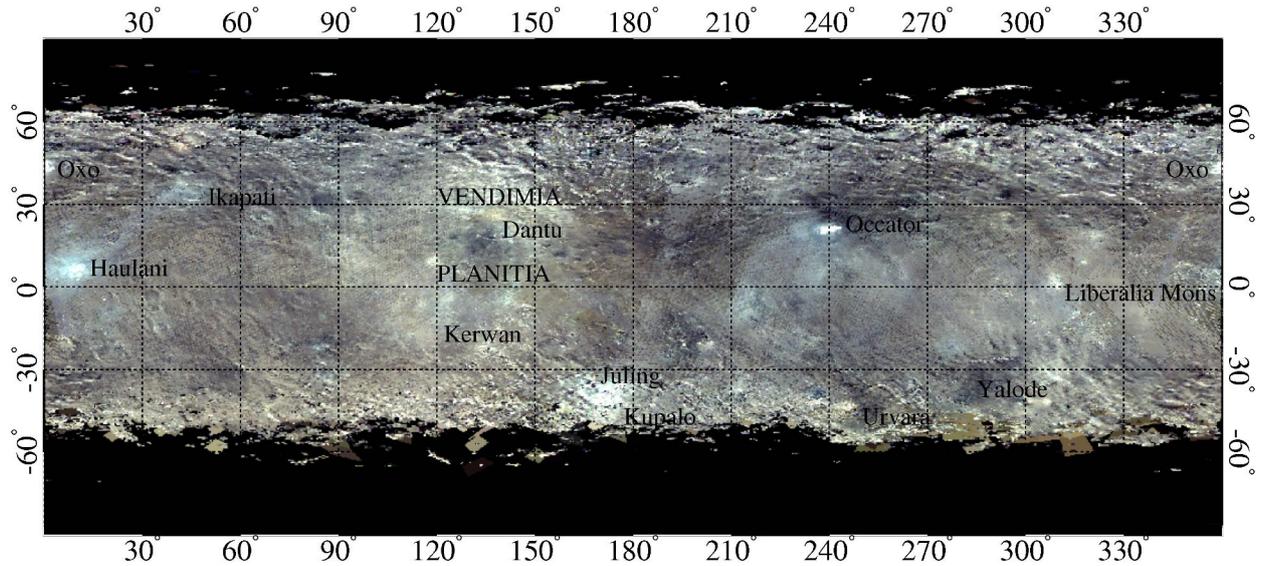

**Fig. 16.** RGB (red = 0.700 $\mu$m, green = 0.550 $\mu$m, and blue = 0.465 $\mu$m) map in the VIS range.

### 4.4. RGB maps

In Fig. 16 an enhanced color map of Ceres' surface in the VIS is shown. A large part of the surface exhibits a reddish spectrum, with bluish features nearby small high-albedo regions such as Haulani, Occator, Oxo, and the Juling-Kupalo area. Conversely, the relatively bright region to the north of Dantu appears redder than the darker central part of Vendimia Planitia. From Haulani an elongated bluish region extends up to Ikapati. This confirms the color distribution shown in the VIS range by Nathues et al. (2015) and Schröder et al. (2017). A similar enhanced color distribution is obtained at IR wavelengths (red = 2.2 $\mu$m, green = 1.8 $\mu$m, and blue = 1.2 $\mu$m, Fig. 17). The main differences with respect to the VIS range are found in the brighter terrains around Vendimia Planitia, which appear bluish in the IR, and on the northeast side of Occator, which shows a reddish spectrum when compared to the blue material surrounding the crater on the southwest part.

## 5. Comparison with other minor bodies

Before the Dawn mission to Vesta and Ceres, several minor bodies, including asteroids and comets, have been closely imaged by remote sensing instruments on board space missions. The observation of these objects under viewing geometries that are not accessible from ground-based facilities enabled comparison of their photometric curves on the basis of their shape (Longobardo et al. 2016) or Hapke parameters. In Table A.1 we report the Hapke parameters at visible-to-infrared wavelengths for a set of asteroids and comets and from previous Ceres' studies, as compared to the parameters derived in this work.

### 5.1. Ceres

The value of the SSA albedo at 0.55 $\mu$m derived from VIR data is much higher than previous measurements by Li et al. (2006), while it is closer to the determination given by Reddy et al. (2015) and Li et al. (2016b). The large difference with the results of Li et al. (2006) can be explained by the different shapes of the SPPF adopted in this work (two terms against a single-term expression) and the limited phase angle range investigated in Li et al. (2006). Substantial differences are also found in the derivation of the $\xi$ parameter, which leads to a more symmetric shape (less back-scattering) of the SPPF than in the result of Li et al. (2006, 2016b) and Reddy et al. (2015). Again, this can be due to the reasons explained above along with the effect of a possible degeneration of the Hapke model parameters (Li et al. 2015): in particular, a more strongly back-scattering SPPF can be compensated for by a reduction of the SSA if the investigated phase angle range is not large enough. A similar explanation can also be valid for the different values of the roughness parameter, which in this work gives $\bar{\theta} = 29°$, while it is estimated to be equal to 44° in Li et al. (2006) and 20° in Li et al. (2016b).

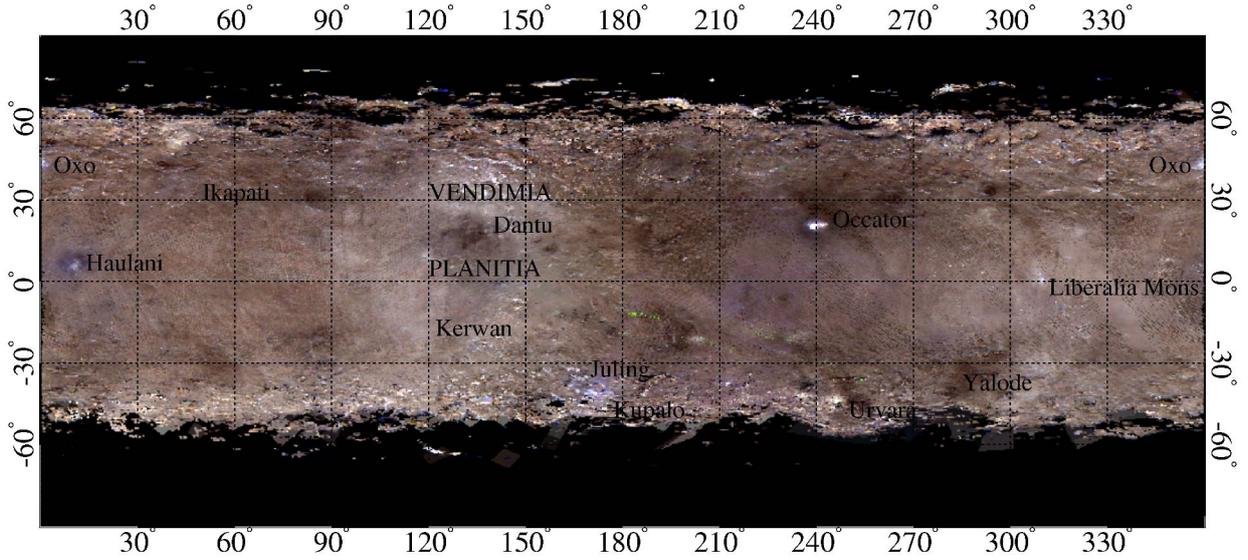

**Fig. 17.** RGB (red = 2.2 $\mu$m, green = 1.8 $\mu$m, and blue = 1.2 $\mu$m) map in the IR range.

### 5.2. Other minor bodies

The comparison with Hapke parameters derived for other minor bodies indicates that Ceres' SSA is intermediate between the dark comets and C-type Mathilde, and the brighter S-type asteroids, while the asymmetry parameter is the largest of the whole dataset. We recall here that the large majority of Hapke models reported in Table A.1 is performed with a single-term Henyey-Greenstein SPPF, which limits the significance of the comparison on $\xi$. The roughness parameter value is among the highest of the distribution, possibly indicating significant subpixel (<1 km scale) roughness.

Because of the coupling between Hapke's parameter and their possible degeneration, a more thorough comparison of the photometric properties of different bodies can be performed by examining quantities that are combinations of the modeled parameters. In this context, the geometric albedo $A_{\rm geo}$ and the integral phase function $\Phi_p(\alpha)$ (Hapke 2012) can give clues on the intrinsic surface albedo properties and the scattering behavior with phase angle. From our modeling we derive $A_{\rm geo} = 0.094 \pm 0.007$, which is in excellent agreement with previous measurements by Reddy et al. (2015) and Li et al. (2006) (Table A.1) and the recent determination by Li et al. (2016a), who reported $0.085 \pm 0.005$. This indicates that Ceres has the darkest surface of the asteroids reported here, with the exception of Mathilde, whose geometric albedo is comparable to those derived for comets. When compared to the average geometric albedo of the different asteroids classes, the albedo for C-type objects represents the closest match (see Table A.2). This is also shown in Fig. 18, where the modeled full-disk reflectance $FDR(\alpha)$, defined as $FDR(\alpha) = A_{\rm geo} \times \Phi_p(\alpha)$, is reported for all the objects of Table A.1: Ceres' curve is positioned above the FDR of comets and C-type Mathilde. In Fig. 19 the integral phase function is shown. In this case, the phase curves can be roughly divided into two classes, with cometary nuclei being more back-scattering, as expected for dark surfaces, and S-type asteroids plus Vesta, which are relatively brighter, being more forward-scattering. Steins stands out from the distribution, with a very shallow integral phase function at small phases. Ceres again shows intermediate properties, surprisingly similar to Lutetia and Mathilde (see box in Fig. 19), characterized by a geometric albedo which is twice and half of the albedo of the dwarf planet, respectively.

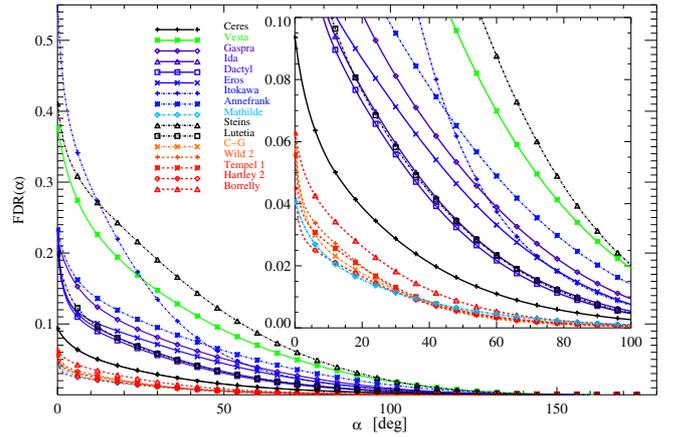

**Fig. 18.** Modeled $FDR(\alpha)$ curves for the objects described in Table A.1 (C-G: Churyumov-Gerasimenko). *Inset panel*: zoom in the $0° < \alpha < 100°$, $0 < FDR(\alpha) < 0.1$ range emphasizing the differences among low-albedo objects.

Asteroid magnitude phase curves from ground-based observations are typically classified by means of the H-G formalism (Bowell et al. 1989; Lagerkvist & Magnusson 1990). To directly compare our results with previous studies, we derived an approximated magnitude phase curve in the Bessel $V$ band (Bessell 1990) $V(1, 1, \alpha)$ (Fig. 20) from the average reflectance computed in 1° phase angle bins $\langle I/F(\lambda)\rangle_\alpha$ by means of the following relation:

$$V(1, 1, \alpha) = m_{\odot,V} - 2.5 \log\left(\frac{\int_V J(\lambda)\langle I/F(\lambda)\rangle_\alpha B_V(\lambda){\rm d}\lambda}{\int_V J(\lambda) B_V(\lambda){\rm d}\lambda}\phi(\alpha)R^2\right),$$
(4)

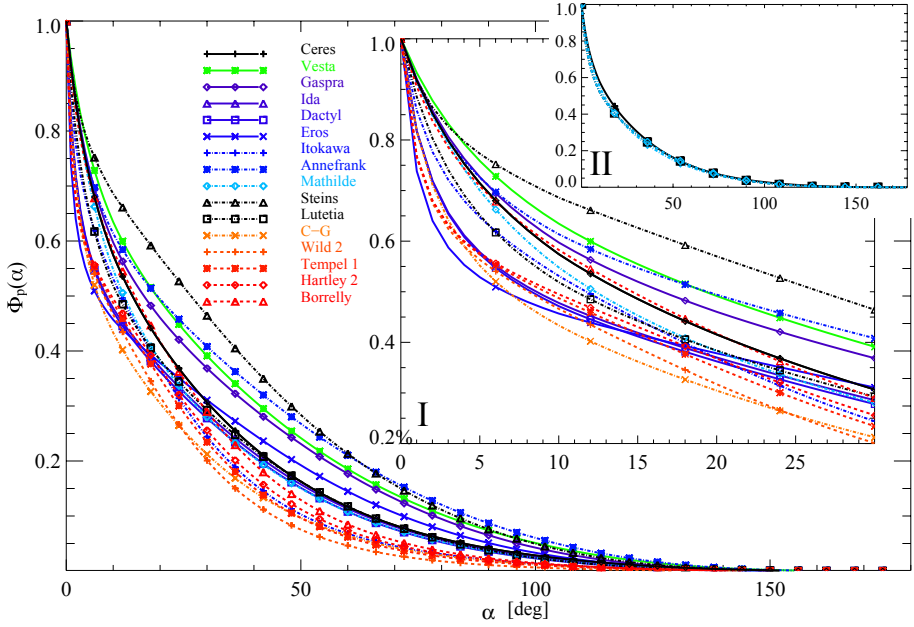

**Fig. 19.** Modeled $\Phi_p(\alpha)$ curves for objects described in Table A.1 (C-G: Churyumov-Gerasimenko). *Inset panel (I)*: zoom in the $0° < \alpha < 30°$ range. *Inset panel (II)*: Ceres' integral phase curve compared to Mathilde and Lutetia: the three curves are superimposed at the scale of the plot.

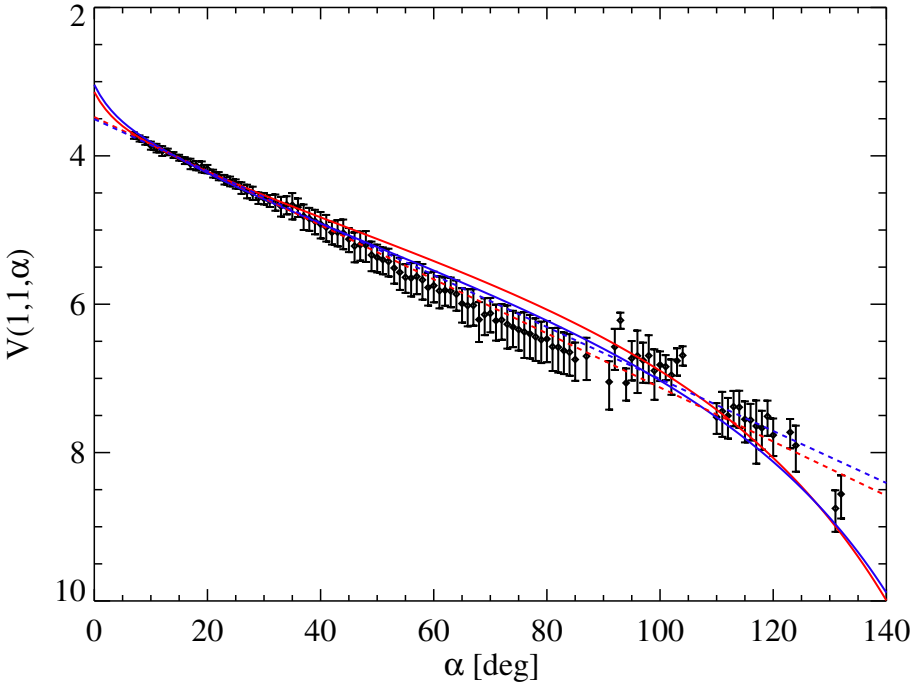

**Fig. 20.** Ceres' magnitude phase curve $V(1, 1, \alpha)$ as derived from VIR disk-resolved observations (diamonds). Error bars are derived from the standard deviation of the spectrally integrated $I/F$ in each phase angle bin. Red curves are derived by fitting the dataset with the H-G formalism (solid line) and linear model (dashed line) for $\alpha \leq 30°$. Blue curves are obtained by fitting the whole phase angle interval.

where $m_{\odot,V}$ is the Sun V-band apparent magnitude, $R$ is Ceres' average radius (470 km) expressed in AU, $J(\lambda)$ is the solar irradiance at 1 AU, $B_V(\lambda)$ is the V-band Bessel filter function, and $\phi(\alpha) = \frac{1+\cos(\alpha)}{2}$ is the fraction of the projected surface that is visible and illuminated. This curve was modeled by applying the H-G expression to observations with phase angles below 30° to be compatible with typical ground-based observation geometries. We obtained $H^a = 3.14 \pm 0.04$ and $G^a = 0.10 \pm 0.04$, while the fit on the full phase angle range gives $H^b = 3.04 \pm 0.02$ and the slope parameter $G^b = 0.02 \pm 0.01$. Similarly, a classical linear model with equation $V(1, 1, \alpha) = V(1, 1, 0°) + \beta\alpha$ (Li et al. 2015) was applied to the magnitude phase curve, and we obtained $V(1, 1, 0°)^a = 3.48 \pm 0.03$, $\beta^a = 0.036 \pm 0.002$ and $V(1, 1, 0°)^b = 3.50 \pm 0.01$, $\beta^b = 0.0350 \pm 0.0004$. These quantities, along with the color indices $V-R$ and $R-I$ and the geometric albedo, are reported in Table A.2 with the corresponding values from previous studies of Ceres and for asteroids and comets. Color indices were directly derived from disk-resolved photometrically corrected observations, but are fully compatible with values obtained from the simulated magnitude phase curves.

The $H$ and $V(1, 1, 0°)$ values derived in this work are lower than results from previous studies (Lagerkvist & Magnusson 1990; Reddy et al. 2015). This difference, in the case of $H$, might be explained by the lack of observations at very low phase angle (<7°), which prevent us from providing an adequate description of the opposition effect surge. The value of the slope parameter $G^a$, derived for $\alpha < 30°$, is compatible with results of Lagerkvist & Magnusson (1990), while a very good agreement is found for $\beta$ among the three different studies. Comparing both $G^a$ and $\beta^a$ with average values of the asteroid classes (Tholen 1984) and individual comets, we find that C-type objects represent the best match, although the phase curve of Ceres is relatively less back-scattering. This is consistent with Ceres having a geometric albedo slightly higher than C-type asteroids. Belskaya & Shevchenko (2000) and Longobardo et al. (2016) showed that the linear slope of asteroid phase curves is related to the surface albedo, with a steeper phase function for darker objects. A comparison of the color indices $V-R$ and $R-I$ points out a similarity with comets 1P/Halley and Hartley 2 and a compatibility with C-type objects, but given the relatively large dispersion of the average color index values of asteroids classes, it is not possible to provide a univocal classification.

## 6. Summary and conclusions

We investigated the spectrophotometric properties of the dwarf planet Ceres by means of Dawn/VIR observations. Disk-resolved images from CSA, RC3, CTS, and CSS orbital phases ($7° < \alpha < 132°$) were analyzed to derive a photometric correction of the dataset by means of Hapke's theory across the VIS-IR (0.465–4.05 $\mu$m) range, and in parallel we assessed the surface physical properties. The derived SSA at 0.55 $\mu$m is $w = 0.14 \pm 0.02$ with an asymmetry factor of the SPPF $\xi = -0.11 \pm 0.08$, while the large-scale roughness parameter is $\bar{\theta} = 29° \pm 6°$. The modeled geometric albedo gives $A_{\rm geo} = 0.094 \pm 0.007$, indicating a dark surface. Phase reddening was measured at VIS and IR wavelengths with values of $4.6 \times 10^{-2}\%$ kÅ$^{-1}$ deg$^{-1}$ and $1.5 \times 10^{-2}\%$ kÅ$^{-1}$ deg$^{-1}$, respectively, while the spectral slopes in the two ranges after photometric correction are $S_{\rm VIS} = 1.1\%$ kÅ$^{-1}$ and $S_{\rm IR} = 0.85\%$ kÅ$^{-1}$.

The whole investigated dataset was photometrically corrected to standard geometry, showing moderate surface albedo variability at a global scale on the order of 9% with a central value of 0.034 at 0.55 $\mu$m and 6% with a central value of 0.032 at 2 $\mu$m. Nonetheless, small-scale structures brighter than the average surface can be recognized, as shown in the albedo maps at 0.55 and 2 $\mu$m. The area exhibiting the highest albedo is the Occator crater, with a reflectance value of 0.12 at 0.55 $\mu$m, although this values is a lower limit of the real reflectance since the brightest spots in the Occator crater are not fully resolved at the dataset resolution (De Sanctis et al. 2016, reported a maximum reflectance of 0.26 from HAMO images). Other notable high-albedo features are represented by the Haulani and Oxo craters. Color maps at VIS and IR wavelength indicate that Ceres is spectrally red. Color variability across the surface is observed, with the brightest areas such as the Occator, Haulani, Oxo, and Juling/Kupalo craters typically exhibiting a bluer spectrum. Interestingly, a few low-albedo regions also appear bluer than the average, like the central part of the Vendimia Planitia region and the terrains surrounding Occator.

A comparison of Ceres' spectrophotometric properties with the properties derived for other minor bodies indicates a phase function behavior that is less strongly back-scattering than comets and average C-type objects ($\beta = 0.036 \pm 0.002$), as well as a slightly higher albedo. Nonetheless, within the usual asteroid taxonomy, C-type appears to be the closest match to the dwarf planet, being also compatible in terms of color indices with $V - R = 0.38 \pm 0.01$ and $R - I = 0.33 \pm 0.02$.

*Acknowledgements.* We thank the following institutions and agencies that supported this work: the Italian Space Agency, the National Aeronautics and Space Administration (NASA, USA) and the Deutsches Zentrum für Luft- und Raumfahrt (DLR, Germany). The VIR was funded and coordinated by the Italian Space Agency and built by SELEX ES, with the scientific leadership of the Institute for Space Astrophysics and Planetology and the Italian National Institute for Astrophysics, and is operated by the Institute for Space Astrophysics and Planetology, Italy. A portion of this work was carried out at the Jet Propulsion Laboratory, California Institute of Technology, USA, under contract to NASA. We also thank the Dawn Mission Operations team and the Framing Camera team.

## Appendix A: Appendix of tables

**Table A.1.** Ceres' Hapke parameters and geometric albedo compared to asteroids and cometary nuclei.

| Target | $SSA$ | $\xi$ | $B_0$ | $h$ | $\bar{\theta}$ | $A_{geo}$ | Phase angle coverage | Asteroid type |
|---|---|---|---|---|---|---|---|---|
| (1)Ceres (this work) | $0.14 \pm 0.02$ | $-0.11 \pm 0.08$ | (1.6) | (0.06) | $29° \pm 6°$ | $0.094 \pm 0.007$ | $7°–132°$ | – |
| (1)Ceres[a] | $0.094–0.11$ | $-0.35 \pm 0.05$ | – | – | $20° \pm 3°$ | $0.085 \pm 0.005$ | $30°–90°$ | – |
| (1)Ceres[b] | $0.070 \pm 0.002$ | $(-0.4)$ | (1.6) | (0.06) | $44° \pm 5°$ | $0.087 \pm 0.003$ | $6.1°–6.2°$ | – |
| (1)Ceres[c] | $0.11$ | $-0.29$ | (1.6) | (0.06) | $(44°)$ | $0.094$ | $0.9°–21.4°$ | – |
| (253)Mathilde[d] | $0.035 \pm 0.006$ | $-0.25 \pm 0.04$ | $3.2 \pm 1.0$ | $0.074 \pm 0.003$ | $19° \pm 5°$ | $0.047 \pm 0.005$ | $40°–138°$ | C |
| (4)Vesta[e] | $0.500$ | $-0.229$ | $1.7$ | $0.07$ | $17.7°$ | $0.38 \pm 0.04$ | $7.7°–80°$ | V |
| (21)Lutetia[f] | $0.25$ | $-0.22$ | $1.58$ | $0.052$ | $24°$ | – | $0°–136.4°$ | M |
| (21)Lutetia[g] | $0.226 \pm 0.002$ | $-0.28 \pm 0.02$ | $1.79 \pm 0.08$ | $0.041 \pm 0.003$ | $28° \pm 1°$ | $0.194 \pm 0.002$ | $0.15°–156°$ | M |
| (2867)Steins[h] | $0.66 \pm 0.02$ | $-0.30 \pm 0.01$ | $0.60 \pm 0.05$ | $0.027 \pm 0.002$ | $28° \pm 1°$ | $0.39 \pm 0.02$ | $0.36°–141°$ | E |
| (951)Gaspra[i] | $0.36 \pm 0.07$ | $-0.18 \pm 0.04$ | $1.63 \pm 0.07$ | $0.06 \pm 0.01$ | $29° \pm 2°$ | $0.22 \pm 0.06$ | $2°–51°$ | S |
| (243)Ida[j] | $0.218^{+0.024}_{-0.005}$ | $-0.33 \pm 0.01$ | $1.53 \pm 0.10$ | $0.02 \pm 0.005$ | $18° \pm 2°$ | $0.206 \pm 0.032$ | $0.6°–109.8°$ | S |
| Dactyl[j] | $0.211^{+0.028}_{-0.010}$ | $-0.33 \pm 0.03$ | (1.53) | (0.02) | $23° \pm 5°$ | $0.198 \pm 0.050$ | $19.5°–47.6°$ | S |
| (433)Eros[k] | $0.33 \pm 0.03$ | $-0.25 \pm 0.02$ | $1.4 \pm 0.1$ | $0.010 \pm 0.004$ | $28° \pm 3°$ | $0.23$ | $0.3°–110°$ | S |
| (25143)Itokawa[l] | $0.36 \pm 0.05$ | $-0.5 \pm 0.1$ | (1.0) | (0.022) | $(20°)$ | $0.53 \pm 0.04$ | $27°–87°$ | S |
| (25143)Itokawa[m] | $0.42 \pm 0.02$ | $-0.35 \pm 0.01$ | $0.87 \pm 0.02$ | $0.01 \pm 0.001$ | $26° \pm 1°$ | $0.33$ | $0.2°–38.4°$ | S |
| (5535)Annefrank[n] | $0.41 \pm 0.05$ | $-0.19 \pm 0.03$ | (1.32) | $0.046 \pm 0.013$ | $(20°)$ | $0.232 \pm 0.038$ | $2.3°–90°$ | S |
| 67P/Churyumov-Gerasimenko[o] | $0.052 \pm 0.013$ | $-0.42$ | – | – | $19°^{+4}_{-9}$ | $0.062 \pm 0.002$ | $1.2°–111.5°$ | – |
| 67P/Churyumov-Gerasimenko[p] | $0.037 \pm 0.002$ | $-0.42 \pm 0.03$ | $1.95 \pm 0.12$ | $0.023 \pm 0.004$ | $15°$ | $0.0589 \pm 0.0034$ | $1.3°–53.9°$ | – |
| 19P/Borrelly[q] | $0.057 \pm 0.009$ | $-0.43 \pm 0.07$ | $1.0$ | $0.039$ | $22° \pm 5°$ | $0.072 \pm 0.020$ | $51°–75°$ | – |
| 103P/Hartley 2[r] | $0.036 \pm 0.006$ | $-0.46 \pm 0.06$ | (1.0) | (0.01) | $15° \pm 10°$ | $0.045 \pm 0.009$ | $79°–93°$ | – |
| 9P/Tempel1[s] | $0.039 \pm 0.005$ | $-0.49 \pm 0.02$ | (1.0) | (0.01) | $16° \pm 8°$ | $0.051 \pm 0.009$ | $63°–117°$ | - |
| 81P/Wild 2[t] | $0.038 \pm 0.04$ | $-0.52 \pm 0.04$ | (1.0) | (0.1) | $27° \pm 5°$ | $0.059 \pm 0.004$ | $11°–100°$ | – |

**Notes.** [a] Li et al. (2016b), clear filter centered at 730 nm, geometric albedo by Li et al. (2016a). [b] Li et al. (2006). [c] Reddy et al. (2015), case 2. [d] Clark et al. (1999), at 0.7 $\mu$m. NEAR data have been combined with telescopic observations at phase angles from 1° to 16°. [e] Li et al. (2013b). [f] Raponi (2015), average values in the 1.1–2.4 $\mu$m wavelength range. These quantities are derived by applying the Hapke (2012) model, including the effect of porosity with $K = 1.18$. [g] Masoumzadeh et al. (2015), at 631.6 nm and 649.2 nm. [h] Spjuth et al. (2012) at 630 nm, Hapke parameters from solution $GS$ 1. [i] Helfenstein et al. (1994) at 0.56 $\mu$m. Values are derived from Earth-based and Galileo mission combined data. [j] Helfenstein et al. (1996), at 0.55 $\mu$m. For Ida, values are derived from Earth-based and Galileo mission combined data. [k] Li et al. (2004), at 0.55 $\mu$m. Near data have been combined with telescopic observations at phase angles below 53°. [l] Lederer et al. (2005), ground-based observations in the V band, solution 4. [m] Kitazato et al. (2008), at 1.57 $\mu$m. Geometric albedo was not reported in the original paper and has been computed here from Hapke parameters. [n] Hillier et al. (2011), at 630 nm. Stardust data have been combined with telescopic observations at phase angles from 2.3° to 18.3°. [o] Ciarniello et al. (2015). [p] Fornasier et al. (2015). [q] Li et al. (2007b), $R$ band. [r] Li et al. (2013a). DIXI observations have been combined with Gemini and HST data at low phase angles. [s] Li et al. (2007a). [t] Li et al. (2009), $R$ band.

**Table A.2.** Spectrophotometric parameters of Ceres compared to asteroids and cometary nuclei.

| Target | $V-R$ | $R-I$ | $H$ | $G$ | $V(1,1,0)$ | $\beta$[mag/°] | $A_{geo}$ |
|---|---|---|---|---|---|---|---|
| Ceres (this work) | $0.38 \pm 0.01$ | $0.33 \pm 0.02$ | $3.14 \pm 0.04^a$ | $0.10 \pm 0.04^a$ | $3.48 \pm 0.03^a$ | $0.036 \pm 0.002^a$ | $0.094 \pm 0.007$ |
|  |  |  | $3.04 \pm 0.02^b$ | $0.02 \pm 0.01^b$ | $3.50 \pm 0.01^b$ | $0.0350 \pm 0.0004^b$ |  |
| Ceres[c] | – | – | $3.24 \pm 0.03$ | $0.076 \pm 0.004$ | $3.58 \pm 0.03$ | $0.038 \pm 0.006$ | $0.099 \pm 0.003$ |
| Ceres[d] | – | – | $3.35 \pm 0.04$ | $0.12 \pm 0.02$ | $(3.69 \pm 0.02)$ | $(0.037 \pm 0.002)$ | – |
| Average C[e] | $0.38 \pm 0.05$ | $0.35 \pm 0.05$ | – | $0.07 \pm 0.01$ | – | $0.043 \pm 0.001$ | $0.06 \pm 0.02$ |
| Average S[e] | $0.49 \pm 0.05$ | $0.41 \pm 0.06$ | – | $0.24 \pm 0.01$ | – | $0.030 \pm 0.001$ | $0.20 \pm 0.07$ |
| Average M[e] | $0.42 \pm 0.04$ | $0.40 \pm 0.05$ | – | $0.20 \pm 0.02$ | – | $0.032 \pm 0.001$ | $0.17 \pm 0.04$ |
| Average E,R[e] | – | – | – | $0.49 \pm 0.02$ | – | $0.024 \pm 0.002$ | $0.52 \pm 0.03^*$ |
| 67P/Churyumov-Gerasimenko[f] | $0.54 \pm 0.05^{**}$ | $0.46 \pm 0.04^{**}$ | – | $-0.09 \pm 0.04$ | – | $0.077 \pm 0.002, 0.041 \pm 0.001$ | $0.062 \pm 0.002$ |
| 1P/Halley[g] | $0.41 \pm 0.03$ | $0.39 \pm 0.06$ | – | – | – | – | $0.04^{+0.02}_{-0.01}$ |
| 103P/Hartley 2[h] | $0.43 \pm 0.04$ | $0.39 \pm 0.05$ | – | – | – | $0.046 \pm 0.002$ | $0.045 \pm 0.009$ |
| 9P/Tempel 1[i] | $0.50 \pm 0.01$ | $0.49 \pm 0.02$ | – | – | – | $0.046 \pm 0.007$ | $0.051 \pm 0.009$ |

**Notes.** [a] Values derived for $\alpha < 30°$. [b] Values derived for the full phase angle range. [c] Reddy et al. (2015). Geometric albedo computed from $H$ parameter. [d] Lagerkvist & Magnusson (1990). $H$ is an average from values by Lagerkvist & Magnusson (1990). The values in parentheses have been calculated by Reddy et al. (2015) from HG models by Lagerkvist & Magnusson (1990). [e] Shevchenko & Lupishko (1998). * This value has been derived for E type asteroids alone. [f] All values from Ciarniello et al. (2015) except ** by Tubiana et al. (2011). $\beta$ values correspond to $\alpha < 15°$ and $\alpha > 15°$, respectively. [g] Thomas & Keller (1989). [h] Li et al. (2013a). [i] Li et al. (2007a).